\documentstyle[12pt]{article}
\def\jour#1#2#3#4#5{{\em#1}:    #2  {\bf#3},        #4,      (#5).}
\def\book#1#2#3#4#5{{\em#1} in: #2, {\tt#3}                  (#5).}
\def\prep#1#2#3#4#5{{\em#1}:    #2  \underline{#3}, {\tt#4}, (#5).}
\def\proc#1#2#3#4#5{{\em#1}:    #2, {\tt#3},        #4,      (#5).}

\newcommand{\be}{\begin{equation}}
\newcommand{\ee}{\end{equation}}
\newcommand{\bea}{\begin{eqnarray}}
\newcommand{\eea}{\end{eqnarray}}
\newcommand{\ba}{\begin{array}{ll}}
\newcommand{\baa}{\begin{array}{lll}}
\newcommand{\ea}{\end{array}}
\newcommand{\bb}{\begin{thebibliography}}
\newcommand{\eb}{\end{thebibliography}}

\newcommand{\lab}[1]{\label{#1}}
\newcommand{\re}[1]{(\ref{#1})}
\newcommand{\Tr}{\mbox{Tr\,}}
\newcommand{\Ds}{\displaystyle}

\begin{document}

\date{May 18, 1994}
\title{ ADDITIONAL DEGREES OF FREEDOM\\ IN SKYRMION MOTION\thanks{
Supported in part by the International Science Foundation
          and Russian Foundation for Fundamental Research 94-02-05837,
          and by the US Department of Energy under Contract
          DE-AC05-85ER40150}}

\author{I.V. Musatov\thanks{Also \em Research Institute of Applied Physics at
Tashkent State University,
Tashkent 700095, Uzbekistan}, V.A. Nikolaev, O.G. Tkachev \\[2.0mm]
{\em Joint Institute for Nuclear Research, 141980 Dubna, Russia} \\[2mm]
{E. Sorace, M. Tarlini} \\[2.0mm]
{\em INFN, Firenze, Italy}
}

\titlepage

\maketitle

\begin{abstract}
We consider the quantization of chiral solitons with baryon number
$B>1$. Classical solitons are obtained within the framework of
a variational approach. From the form of the soliton solution
it can be seen
that besides the group of symmetry describing transformations of the
configuration as a whole there are additional symmetries corresponding
to internal transformations.
Taking into account the additional degrees of freedom leads to some sort
of spin alignment for light nuclei and gives constraints on their spectra.
\end{abstract}

\newpage

\section{Introduction}

The considerable recent interest in the Skyrme model \cite{Skyrme1}
as a possible theory of strongly interacting particles is a consequence
of the hope
that meson effective Lagrangians can bridge the gulf between quantum
chromodynamics (QCD) and the known theory of nuclear structure.

Although everyone believes that physics of any nucleus is also
described by the QCD Lagrangian, no one has been able to obtain the basic
properties of nuclei in terms of quark and gluon fields.
It is very difficult to analyze the dynamics
of the quark and gluon fields in low-energy
quantum chromodynamics because of the large coupling constant.

 Searching for a small parameter in QCD, 't Hooft proposed
the idea of considering QCD with a large (tending to infinity) number
of colors $N_c$. Later, Witten showed that if the limit $N_C\to\infty$
exists, then QCD is a theory of effective meson fields with local
interactions with a coupling constants of order of $1/N_c$.
Moreover, in this limit the baryon masses
prove to be of an order of $N_c$, while the number of colors completely
drops out  from the equations determining the size and structure of the
baryons \cite{WittenNuc}.

It is well known that nonlinear theories can have solutions corresponding
to localized objects of finite size --- solitons \cite{Faddeev}-- with the
analogous dependence of the size on the coupling constant.
Therefore, Witten's result
leads to the description of baryons as solitons of an effective meson
theory.
This picture does not require any further reference to the quark origin of the
effective Lagrangian. A theory of just this type was proposed
by Skyrme in 1961-1962 \cite{Skyrme1}.

Nonlinear chiral theories naturally lead to soliton sectors. Already at the
classical level, chiral solitons are very similar to hadrons. They carry a
definite, rigorously conserved topological charge. This localized charge is a
good candidate for the baryon number. Chiral solitons are extended,
strongly interacting objects. They have very large mass compared with
the masses of the fields involved in the Lagrangian.

These features plus  a rich spectrum of generated states make
chiral dynamics a very attractive theory for low-energy phenomena in
strong interaction physics.

Restricting ourselves to the simplest model of this type - the Skyrme
model -,
we probably cannot hope for good quantitative agreement with the experimental
data, but we can obtain a qualitatively good description of the fundamental
regularities characterizing a system of strongly interacting particles which
would support the idea that baryons are solitons of the effective meson
Lagrangian.

The Skyrme model gives us a straightforward way
for constructing a system with an arbitrary baryon charge. We have to look for
solitons of classical fields with corresponding topological charge and
then to quantize solitonic degrees of freedom to obtain an object with nuclear
quantum numbers.

Recently a specific variational ansatz was proposed independently
in \cite{Nikolaev3} and \cite{Tarlini}.
This ansatz obeys the symmetry conditions formulated
in \cite{Verbaarshot}, \cite{Manton} and, being very simple, gives the
possibility to do one more step in the analytical  study
of the problem and to
take into account vibrational modes, for instance, the monopole one,
in a simple way. This analysis gives a natural explanation of the origin
of the ansatz used earlier in \cite{Weigel} and also gives some new solutions.

To obtain quantum spectra of multibaryon, one has to perform  the
quantization of pion field around the multisoliton classic
field configuration. It is well known that the Lagrangian describing
the quantum pion field contains zero modes, which are determined
by the symmetry group of the classical soliton solution.
The zero modes should be treated in a special way. The most
convenient method is to introduce corresponding collective coordinates.
This leads to the interpretation of the soliton as
a quantum particle moving in the collective coordinate space.
As  we will show, the multisoliton solutions obtained with
the anzatz \cite{Nikolaev3,Tarlini} possess additional internal group of
symmetry.
As a consequence, new restrictions on the spectra of multibaryons
arise.

\section{Ansatz and Solutions for Static Equations}

     Here we follow  the paper \cite{Nikolaev5}
(see also \cite{Nikolaev7}).
 For a variational treatment we use the chiral field $U$
\begin{eqnarray}
U(\vec r ) = \cos F(r) + i(\vec {\tau} \cdot \vec N )\ \sin F(r) .
\label{e1}
\end{eqnarray}
with the following assumption about the configuration of the isotopic vector
field $\vec N$ :
\begin{eqnarray}
\vec N = \{ \cos(\Phi (\phi ,\theta ))  \sin(T(\theta )),
\ \sin(\Phi (\phi ,\theta ))  \sin(T(\theta )),\ \cos(T(\theta )) \} .
\label{e2}
\end{eqnarray}
Here $\Phi (\phi,\theta),\ T(\theta)$ are some
arbitrary functions of angles $(\theta ,\phi)$ of the vector $\vec r$
in the spherical coordinate system.
For simplicity and taking in account the qualitative content of
the numerical analysis,  we dropped
the dependence (assumed in \cite{Tarlini}) of $T$ on the radial
variable $r$ and dependence of $F$ on the angular variable $\theta$.

     Let us consider the Lagrangian density ${\cal L}$ for the stationary
solution:
\begin{eqnarray}
{{\cal L}} = {F_\pi^2\over 16}  Tr(L_k L_k)+{1\over 32e^2}
Tr\Bigl[L_k,L_i\Bigr ]^2 .\label{e3}
\end{eqnarray}
Here $L_k= U^+{\partial}_kU$ are the left currents.

     Variation of the functional $L=\int {\cal L}d\vec r$ with respect to
$\Phi$ leads to an equation which has a solution of the type
\begin{eqnarray}
\Phi(\theta ,\phi) = k(\theta) \phi+c(\theta)\nonumber
\end{eqnarray}
with a constraint
\begin{eqnarray}
{\partial\over\partial\theta}\left[\sin^2T(\theta)  \sin\theta
{\partial \Phi(\theta,\phi)\over\partial\theta}\right] = 0. \label{e8}
\end{eqnarray}
It is easily seen from eq. (\ref{e8}) (see also \cite{Nikolaev5})
that functions $k(\theta)$ and $c(\theta)$ may be piecewise constant
functions (step functions):
\begin{eqnarray}
\Phi(\theta ,\phi) = \left\{
  \matrix{
    \displaystyle{k^{(1)} \phi+\rho^{(1)}\ ,
          \ \ {\rm for}\ 0\leq\theta < \theta_1\ ,}\hfill\cr\cr
    \displaystyle{k^{(2)} \phi+\rho^{(2)}\ ,
          \ \ {\rm for}\ \theta_1\leq\theta < \theta_2\ ,}\hfill\cr\cr
    \displaystyle{\ .\ \ \ \ .\ \ \ \ .}\hfill \cr\cr
    \displaystyle{k^{(n)} \phi+\rho^{(n)}\ ,
           \ \ {\rm for}\ \theta_{n-1}\leq\theta < \pi\ .}\hfill
  }\right.\nonumber
\end{eqnarray}

Moreover, $k^{(m)}$ must be integer in any region
$\theta_m\leq\theta\leq\theta_{m+1}$,
where $\theta_m$, $\theta_{m+1}$ are successive points of discontinuity
of $\partial \Phi_i(\theta ,\phi) / \partial \theta$.
The positions of these are the points determined by the condition
\begin{eqnarray}
T(\theta_m) = m \pi\ ,\ \ \ T(\pi) = n \pi \label{e9}
\end{eqnarray}
with integer $m$, as follows from eq.(\ref{e8}).

The soliton mass is given by a functional which can be represented as a
sum of contributions from different $\theta$ - regions.
The functions $F (x)$ and $T (\theta)$ have to obey the equations
derived in \cite{Nikolaev5}, in each $\theta$ - region with given
number $k^{(m)}$.

\section{The Number of Zero Modes}

Let us consider the quantization of the static multibaryon
configuration (\ref{e2}),(\ref{e8}). This procedure implies that the pion
field is represented in the form of a superposition of the background
classical field $\varphi_c (\vec x)$ plus small (quantum) fluctuations around
it:

\be
\varphi(\vec x, t) = \varphi_{c}(\vec x) + \pi(\vec x,t) .
\ee
Then action for quantum pion field can be expanded into series
in $\varphi_q$:
\be
S(\varphi) = S_0(\varphi_c) + {1\over 2} \int dx dy \;\; \pi^a(x)
         \left(\left.
         {\delta^2 S(\varphi) \over \delta \varphi^a(x) \delta \varphi^b(y)}
         \right|_{\varphi=\varphi_c} \right)   \pi^b(y) + \dots,
\lab{QA}
\ee
linear term vanishes as a consequence of equations of motion.

A well--known problem arises due to the zero modes
in  \re{QA}.
In   terms of the path integral quantization it means
that some of the integrations in the functional space are non-Gaussian and
should be carried out with a specific procedure
(rather than in the saddle-point approximation).

First question is about the number of zero modes.
The reason for these zero modes is that a soliton solution breaks
explicitly some of the symmetries of the initial Lagrangian, and each
mode restores relevant symmetry of the partition function. So,
the usual way to treat them is to extract the volume of the symmetry group,
in particular, introducing a set of time-dependent collective coordinates
$\alpha(t)$.
Thus the measure in the path integral can be modified,
by inserting the Faddeev-Popov unity, into the form
\be
Z= \int {\cal D} \pi e^{iS(\varphi_c; \pi)}
   = \int {\cal D} \{\alpha\} \int {\cal D} \pi' e^{iS(\varphi_c,
\{\alpha\}; \pi)},
\ee
where prime denotes that zero modes are excluded from the path
integral measure over the pion field.

The collective coordinates can be chosen
as the parameters of the soliton solution
$\varphi_c(\vec x;t) = \varphi_c(\vec x; \{\alpha(t)\})$, the classical action
$S_0(\varphi_c;\{\alpha\})$ being in fact independent on $\alpha$'s.
First of all, the parameters are those defining the global
transformations of a soliton, which are the coordinate of center $\vec X$ and
the matrices of orientation in configurational and iso-spin spaces
$R$ and $I$ respectively:
\be
U(\vec x; t)=
              e^{-i \vec P \vec X}
              e^{i TI }
              e^{-i SR}\cdot
              U_0(\vec x) =
              \exp \left\{ i \tau^i I^{ij}(t)
              N^j\left( R^{(-1)}_{kl}(t) x_k\right)\,
              F(|\vec x - \vec X|)\right\}
\lab{coord}
\ee
Here we denote generators of the rotations in space and
iso-space  as $S$ and $T$.

In general, the multisoliton field configuration  (\re{e2},\re{e8})
allows for wider
group of symmetry due to specific form of the ansatz. The action can be
seen to be independent on the parameters
{\bf $\rho^{(i)}$} which define the orientation of
$i$-th sector $\theta \in [\theta^{(i-1)},\theta^{(i)}]$ in the $xy$--plane.

However, not all of the parameters we have introduced are in fact
independent. To see this, let us represent the matrices $I$ and
 $R$  (and the relevant generators) as a composition of
the two parts:
\be
I= I_\perp   I_3,\ \ \ \ R= R_\perp   R_3
\lab{I3}
\ee
where $R_3$ ($I_3$) describes the rotation around the z-- (third) axis in
space (isospace),
and $R_\perp$ ($I_\perp$) describes the rotation around
an axis lying in the xy-- (12) plane.
Then, note that instead of the parameters $\rho^{(i)}$ the set of
matrices
$R^{(i)}(\theta)$ can be introduced, so that
\be
R_3^{(i)}(\theta)=\left\{ \begin{array}{ll}
            R_3 (\rho^{(i)}), &\theta\in [\theta^{(i)},\theta^{(i+1)}],\\
            1,                     &\mbox{otherwise} .
            \end{array} \right.
\lab{Ii}
\ee

Obviously, $R_3^{(i)}  R_3^{(j)}=R_3^{(j)}  R_3^{(i)}$. Let us
define, in the analogy with
\re{Ii}, the set of operators $S_3^{(i)}(\theta)$ and
$T_3^{(i)}(\theta)$, which
rotate the $i$-th sectors around the z-- (third) axis independently.
It is easy to check that
\be
S_3^{(i)} - k^{(i)}   T_3^{(i)} =0,
\lab{TpluskM}
\ee
and
\be
S_3=\sum
\limits_{i=1}^n S_3^{(i)}= \sum \limits_{i=1}^n k^{(i)}
T_3^{(i)}, \ \ \
T_3=\sum
\limits_{i=1}^n T_3^{(i)}.
\lab{T3M3}
\ee

As a result, we see that independent operators of the space and
iso--space rotations can be chosen as $S_\perp$, $T_\perp$
and the set of $T_3^{(i)}$
(one can equivalently choose another operator
basis of the same dimension). So, independent collective coordinates
are $R_\perp$, $I_\perp$ and the set of $\rho^{(i)}$.

\section{Lagrangian in the Collective Coordinate Variables}

We want to find the spectrum of low-lying quantum states of a
multibaryon which corresponds to the classical multisoliton field
configurations Eqs.(\ref{e2},\ref{e8}). This can be performed
by means of the canonical quantization method.

For our purpose the zero modes corresponding to the rotational symmetries
seem the most interesting, since they
determine the rotational spectrum structure of low-lying multibaryon
states.
Therefore, we restrict ourselves here to consideration of the zero modes.

The natural way to proceed is to rewrite the Lagrangian in
terms of the independent collective coordinates and their time derivatives
and to derive the Hamiltonian.
However, it is more instructive
to keep the overfull set of the parameters
$R$, $I$ and $\rho^{(i)}$, i.e. not to separate out
the overall rotation and iso-- rotation around the z-(third) axis. We will
obtain the constraints (\ref{TpluskM}) again at the end of the calculations.

It is convenient to define the angular velocities by

\be
R_{ik}^{-1} \dot R_{kj} =\epsilon_{ijl}\Omega_l, \ \ \ \
\dot I_{kj} I_{ik}^{-1} =\epsilon_{ijl}\omega_l.
\ee

Inserting Eqs. (\ref{coord}), (\ref{I3}) and (\ref{Ii}) into the Lagrangian,
we obtain
\be
{\rm L}=-M+{F_\pi^2 \over 16}  \int \Tr(L_0 L_0) d^3 x
    +{1\over 16e^2}  \int \Tr [L_0, L_i]^2 d^3 x = -M+{\rm L}'
\ee
and
\be
{\rm L}'={1\over 2} \left\{
   {\vec \Omega_\perp}^2 Q_S + {\vec \omega_\perp}^2 Q_T
   + \sum \limits_{i=1}^N
       \left( \omega_3 k^{(i)} + \dot \rho^{(i)} + \Omega_3 \right)^2 C^{(i)}
   + 2  (\Omega_1\omega_1 K_1
      + \Omega_2\omega_2 K_2)
   \right\}.
\lab{Lagr}
\ee
Here ${\vec \Omega_\perp}^2 = \Omega_1^2 + \Omega_2^2$,
${\vec \omega_\perp}^2 = \omega_1^2 + \omega_2^2$ and
$\vec \omega_\perp   \vec \Omega_\perp =
  \omega_1 \Omega_1 + \omega_2 \Omega_2$, M is the classical
energy of the soliton.
Each of the quantities $Q_{S,T}$ and $K_i$ in the above equation may be
considered as a sum of independent contributions from different sectors
$\theta \in [\theta_{i-1},\theta_{i}]$:
\be
Q_{S,T} = \sum \limits_{i=1}^N Q_{S,T}^{(i)},\ \ \ \
K_{1,2} = \sum \limits_{i=1}^N K_{1,2}^{(i)}.
\ee

Explicit expressions for all the parameters in eq. (\ref{Lagr}) are given in
Appendix A.

$K_{1,2}$ do not vanish only if there is at least one sector
with $|k|=1$.
We will consider multisolitons with all $k^{(i)}$ positive. In this case
$K_1=K_2=K$ and the sum in the last parenthesis gives
$K \vec\Omega_\perp   \vec\omega_\perp$, so
the system is a symmetrical rotator.

\section{The Hamiltonian for a Quantized Multibaryon}

Let us introduce the canonical momenta conjugated to each of the collective
coordinates

\be
T_m = {\delta L \over \delta \omega_m},\ \ \ \
S_m = {\delta L \over \delta \Omega_m},\ \ \ \
W^{(i)} = {\delta L \over \delta \dot\rho^{(i)}}.
\ee

After the canonical transformation one arrives at the expression for
the Hamiltonian

\be
H = M + { {\vec S}^2 \over 2 {Q'}_S } + { {\vec T}^2 \over 2 {Q'}_T }
  - {S_3^2 \over 2 {Q'}_S } - {T_3^2 \over 2 {Q'}_T }
  + {\vec S   \vec T \over Q_{ST} }
  + \sum\limits_{i=1}^N { {W^{(i)}}^2 \over 2 C^{(i)} },
\lab{Ham}
\ee
If no sectors with $|k^{(i)}|=1$ are presented the new parameters
${Q'}_S$ and ${Q'}_T$ in Eq.\ref{Ham} coincide with
$Q_S$ and $Q_T$ respectively and ${Q_{ST}} \to \infty$.
Otherwise,
\be
{Q'}_S= Q_S - {K^2 \over Q_T}, \ \ \
{Q'}_T= Q_T - {K^2 \over Q_S}, \ \ \
Q_{ST}= -K + {Q_s Q_T \over K}.
\ee

The operators $T_3$ and $S_3$ are not independent and are
related to the set of $W^{(i)}$ via the constraints
\be\baa
T_3 = &\Ds \sum\limits_{i=1}^N W^{(i)},  & T_3^{(i)}= W^{(i)} \\
S_3 = &\Ds \sum\limits_{i=1}^N k^{(i)}   W^{(i)},
        & S^{(i)}= k^{(i)}   W^{(i)} ,
\label{T3S3}
\ea\ee
which are consistent with Eqs. (\ref{TpluskM}) and (\ref{T3M3}).

Note that these relations hold only in the internal frame.

\section{Quantum spectra of multisolitons and numerical results}

We want to construct quantum states of a multisoliton as compositions
of quantum states of individual sectors (regions $[\theta_{(i-1)},
\theta_{(i)}]$), which have definite spin and isospin quantum numbers:
\begin{equation}
| S^{(i)}, T^{(i)}, S^{(i)}_3=k^{(i)} T^{(i)}_3 \rangle.
\label{secstate}
\end{equation}
To this end, we define the most general composition and then step
by step apply the restrictions which follow from the form of
the Hamiltonian and from the rotational symmetry.

Note, that this problem is different from the standard
problem of constructing quantum states of a system of
spinning particles. It is due to the specific form of our
Hamiltonian, which possesses definite quantum numbers
not only for total spin and isospin with their third
components but also for the operators $T^{(i)}_3$
(and, as a consequence, for
$S^{(i)}_3$) for each of the sectors.

Global spin and isospin rotational symmetry
dictates that a multisoliton quantum state should have the form
of a linear combination
\begin{equation}
\Psi (S,T,S_3,T_3,T^{(i)}_3)= \sum \limits_{{T^{(i)}}, {T^{(i)}_3}'}
                     c_{T^{(i)}, {T^{(i)}_3}'}
                     \psi (S,T,S_3,T_3,{T^{(i)}, {T^{(i)}_3}'})
\label{Psi}
\end{equation}
of the expressions
\begin{equation}
\psi (S,T,S_3,T_3,{T^{(i)}, T^{(i)}_3}) =
\sum \limits_{T^{(i)}_3=-T^{(i)}}^{T^{(i)}}
        C^{S,S_3}_{\{S^{(i)}_3 \}} C^{T,T_3}_{\{T^{(i)}_3 \}}
 \prod \limits_{}^{}
 | S^{(i)}, T^{(i)}, S^{(i)}_3=k^{(i)} T^{(i)}_3 \rangle.
\label{psi}
\end{equation}
Here $C$ are the $3nJ$ --symbols, and we used the relation (\ref{T3S3}).

For the sake of simplicity,
we will illustrate the general idea of our calculation for the case of the
multisoliton configuration with two sectors and dismiss
the spin quantum numbers $S,S_3$; the calculation can be easily
extended to a multisoliton with arbitrary number
of sectors and for the full set of the variables.

First of all, from the requirement that the
multisoliton state must be an eigenstate
of the operators
$\hat T_3, \hat T^{(1)}_3, \hat T^{(2)}_3$
we see that the sum (\ref{Psi},\ref{psi})
contains only one term, with
\begin{equation}
T_3=T^{(1)}_3+T^{(2)}_3,\quad\quad T^{(i)} = |T^{(i)}_3|.
\label{totalT3}
\end{equation}
Furthermore, since
$\hat T = \hat T^{(1)} + \hat T^{(2)}$,
for the operator of the total isospin squared
we have:
\begin{equation}
{\hat T}^2
     = {\hat {T^{(1)}}}^2 + {\hat {T^{(2)}}}^2 +
         2   \hat T^{(1)}_3   \hat T^{(2)}_3 +
                 \hat T^{(1)}_+   \hat T^{(2)}_- +
                 \hat T^{(1)}_-   \hat T^{(2)}_+
\label{totalT}
\end{equation}
On the other hand,
\begin{equation}
{\hat T}^2  |T,T_3;T^{(1)}_3,T^{(2)}_3 \rangle =
T(T+1)      |T,T_3;T^{(1)}_3,T^{(2)}_3 \rangle
\label{Tsqueared}
\end{equation}
and $T=T^{(1)}+T^{(2)}$, which
is consistent with
(\ref{totalT3}) and (\ref{totalT}) only if the two
last terms in (\ref{totalT}) vanish on
the state vector from (\ref{Tsqueared}). Together with Eq.(26) which states
that $T_3^{(i)}$ have their maximum possible values, it
leads to the conclusion that
both $T^{(1)}_3,T^{(2)}_3$ have the same sign.

As a result, we see that the multisoliton quantum
state
$|S,T,S_3,T_3;T^{(i)} \rangle$
has the form of a product of the sector's
quantum states (\ref{secstate}) satisfying the relations
\begin{eqnarray}
&&T= \sum T^{(i)}, \ \
  S= \sum S^{(i)}, \ \
  T_3= \sum T^{(i)}_3, \ \
  S_3= \sum S^{(i)}_3, \\
&&T^{(i)}_3 = +T^{(i)} \ \ ({\rm or} \ T^{(i)}_3 = -T^{(i)} ), \\
&&S^{(i)}_3= k^{(i)}   T^{(i)}_3, \ \ S^{(i)}=|S^{(i)}_3|.
\end{eqnarray}

Substituting these relations into the Hamiltonian (\ref{Ham})
gives the energy of a soliton
\begin{eqnarray}
E = M + {S\over 2Q_S} + {T\over 2Q_T} + {ST\over Q_{ST}}
       + \sum\limits_j {(T_3^{(j)})^2\over 2C^{(j)}}
\label{7}\end{eqnarray}
and the constraint on its spin and isospin quantum numbers
\begin{eqnarray}
T = {\rm max}\left|\sum_j T_3^{(j)}\right|.
\label{8}\end{eqnarray}

Note, that if there are no sectors with $k^{(i)}=1$, the forth term
in (\ref{7}) vanishes and energy of the soliton is linear in total
spin and total isospin.

Corresponding calculations of the rotational energies
for the solitons with baryon number three have been worked out.
In \cite{Nikolaev5} it was shown that the toroidal configuration
($L=1$, $k =\{3\}$) can not have t and $^3$He quantum numbers.
In fact, the corresponding quantum numbers are $T=1/2$, $S=3/2$
but not $T=1/2$, $S=1/2$ as it has to be for t and $^3$He.
It is easy to see from the last formulas that only non-toroidal
configuration ($L=2$, $k =\{1,2\}$) can have
correct quantum numbers after quantization.
Their masses are equal to each other (possible coulomb mass
differences are neglected).
We have to note here that L.Carson has considered the minimal-energy solution
with  $B=3$ of the SU(2) Skyrme model with tetrahedral symmetry and
shown that this discrete symmetry ensures that the $J^\pi={1\over2}^+$
isodublet nucleous ($^3$He, $^3$H) emerges as the unique ground state when
its isospin and rotational zero modes are quantized \cite{CarsonX}.

  From equation (\ref{7}) one obtains that the rotational motion energy
is about $23.5\ MeV$. The classical part of the mass $M$ in eq.(\ref{7})
is $2987\ MeV$.
The values of the constants $F_\pi=109.45\ MeV$ and $e=4.138$
which have been used in our calculations
correspond to the values at which the smallest masses of the
solitons with $B=4$ and $B=12$  coincide with the masses
of the $^4$He and $^{12}$C nuclei \cite{Nikolaev10}.
It is evident that the adiabatic rotation motion approximation
is more convenient for nuclei than for nucleon.

\section{Conclusion}

The quantization procedure including the additional new zero modes
for the non-toroidal soliton configurations has been developed.
The obtained  effective Hamiltonian leads to new formulas for eigenvalue
spectra of the quantum solitons due to the additional constraints we have
obtained for the quantum numbers of considered solitons.
The non-toroidal solitons ($L=2$, $k =\{1,2\}$) have
correct quantum numbers of t and $^3$He after quantization in contrast to
the pure toroidal configurations. We have to note here that it is only
taking into account the additional zero modes that leads to this successful
picture.

\section{Acknowledgments}

One of us (I.M.) is grateful for the warm hospitality to the Theory
Group of CEBAF where the work was completed.

\newpage
\appendix
\section{Formulas for the Moments of Inertia}

 Here we list the
 explicit expressions for the parameters in Eq.(\ref{Lagr}).
 It is customary to use the dimensionless variable
$x=F_\pi e r$ instead of r.

\be \ba
Q_T^{(i)} = &\Ds {\pi\over F_\pi e^3}
      \int\limits_{\theta_{i-1}}^{\theta_i} \sin \theta d\theta
      \int\limits_0^\infty  x^2 dx
      \left\{ -{\sin^4 F\over x^2}
      \left( k_i^2 {\sin^2 T \over \sin^2\theta} \cos^2 T +(T')^2 \right)
        \right. \\ & \Ds \left.
    + \sin^2 F   \left[
      {1\over 4}+
      (F')^2 + \left({k_i^2 \sin^2 T \over \sin^2 \theta} + (T')^2  \right)
      {\sin^2 F \over x^2} \right]
              (1 + \cos^2 T )\right\},
\ea
\ee

\be \ba
Q_S^{(i)} = &\Ds {\pi\over F_\pi e^3}
      \int\limits_{\theta_{i-1}}^{\theta_i} \sin \theta d\theta
      \int\limits_0^\infty  x^2 dx
      \left\{ -{\sin^4 F\over x^2}
      \left( k_i^4 {\sin^4 T \over \sin^4\theta} \cos^2 T + (T')^4 \right)
        \right. \\ & \Ds \left.
    + \sin^2 F   \left[
      {1\over 4}+
      (F')^2 + \left({k_i^2 \sin^2 T \over \sin^2 \theta} + (T')^2  \right)
      {\sin^2 F \over x^2} \right]
        \right. \\[5mm] & \Ds \left. \ \ \
        \left( k_i^2 {\sin^2 T \over \sin^2\theta} \cos^2 \theta
                   + (T')^2 \right) \right\},
\ea
\ee

\be\ba
K_1^{(i)} =  &\Ds
      \delta_{k_i,1} {\pi\over F_\pi e^3}
      \int\limits_{\theta_{i-1}}^{\theta_i} \sin \theta d\theta
      \int\limits_0^\infty  x^2 dx
      \left\{ -{\sin^4 F\over x^2}
        \right. \\ & \Ds \left.
      + \sin^2 F   \left[
      {1\over 4}+
      (F')^2 + \left({k_i^2 \sin^2 T \over \sin^2 \theta} + (T')^2  \right)
      {\sin^2 F \over x^2} \right] \right\}
       \\[4mm] & \Ds \ \ \
        \left( T' \sin \theta
      - \sin T \ \cos T {\cos \theta \over \sin \theta} \right),
      \\[5mm]
K_2^{(i)}  = &\Ds k_i   K_1^{(i)}.
\ea\ee

\be \ba
C^{(i)} = &\Ds {\pi\over F_\pi e^3}
      \int\limits_{\theta_{i-1}}^{\theta_i} \sin \theta d\theta
      \int\limits_0^\infty  x^2 dx
      \left\{ -{\sin^4 F\over x^2}
       k_i^2 \sin^2 T
           \right. \\ & \Ds \left.
    + \sin^2 F   \left[
      {1\over 4}+
      (F')^2 + \left({k_i^2 \sin^2 T \over \sin^2 \theta} + (T')^2  \right)
      {\sin^2 F \over x^2} \right] \right\},
\ea
\ee

\noindent
where  $T^\prime={\partial T(\theta) \over \partial\theta}$,
$F^\prime={\partial F(x) \over\partial x}$.

\end{document}